\documentclass[onecolumn,showpacs,preprintnumbers,amsmath,amssymb]{revtex4}
\usepackage{graphicx}
\usepackage{dcolumn}
\usepackage{bm}

\newcommand{\bef}{\begin{figure}}
\newcommand{\eef}{\end{figure}}

\newcommand{\be}{\begin{equation}}
\newcommand{\ee}{\end{equation}}
\newcommand{\bea}{\begin{eqnarray}}
\newcommand{\eea}{\end{eqnarray}}

\begin{document}

\title{An experimental review on heavy flavor $v_{2}$ in heavy-ion collision}



\author{Md. Nasim, Roli Esha, Huan Zhong Huang
}
\affiliation{Department of Physics and Astronomy, UCLA, CA-90095, USA.
 }

\begin{abstract}
For over a decade now, the primary purpose of relativistic heavy--ion collisions at the
Relativistic Heavy Ion Collider (RHIC) and the Large Hadron Collider
(LHC) has been to study the properties of QCD matter under extreme
conditions -- high temperature and high density.  The heavy--ion experiments at both RHIC and LHC have recorded a
wealth of data in p+p, p+Pb, d+Au, Cu+Cu, Cu+Au, Au+Au,
Pb+Pb and U+U collisions at energies ranging from
$\sqrt{s_{NN}}$ = 7.7 GeV to  7 TeV.  Heavy quarks are considered good probe to study the QCD matter created in relativistic
collisions due to their very large mass and other unique properties. A precise measurement of various properties of heavy flavor  hadrons provides
an insight into the fundamental properties of the hot and dense medium created in these nucleus--nucleus collisions, such as
transport coefficient and thermalization and hadronization
mechanisms. The main focus of this paper is to present  a review on
the measurements of azimuthal anisotropy of heavy flavor hadrons and to
outline the scientific opportunities in this sector
due to future detector upgrade.
We will mainly discuss the elliptic flow of open charmed meson
($D$-meson), $J/\psi$ and leptons from heavy flavor decay at RHIC and LHC energy. 
\end{abstract}
\maketitle

\section{Introduction}
In the standard model of particle physics, the strong force is described by the theory of Quantum Chromodynamics (QCD). At ordinary temperatures or densities this force just confines the quarks into hadrons. At sufficiently high temperature and/or high baryon density, Lattice QCD predicts a transition form hadronic matter to deconfined partonic matter~\cite{qcd_1,qcd_2,qcd_3,qcd_4}. Phase diagram of QCD is full of puzzles and surprises. Experimentally it is possible to create very high temperature and high dense states of nuclear matter, by colliding two heavy nuclei at ultra-relativistic speed, which would contain asymptotically free quarks and gluons. The Relativistic Heavy Ion Collider (RHIC) and the Large Hadron Collider (LHC) was  built to study the properties of  Quark-Gluon Plasma (a state of matter believed to exist  just after the Big Bang) produces during collision of heavy ions at ultra-relativistic speed. The life-time of Quark-Gluon Plasma (QGP) is very short ($\sim$5-10 fm/c), hence direct detection of QGP is not possible. Therefore, one has to rely on indirect measurement using suitable probe.  Heavy quarks production  in relativistic heavy-ion collisions provides unique probes of QGP.\\
In relativistic heavy ion collisions, heavy quarks (c, b) are produced on a short time scale ($\sim$0.08 fm/c for $c\bar{c}$ production) in the initial hard partonic scatterings during the early stages of nucleus-nucleus collisions~\cite{pbm}. The probability of thermal production of heavy-quark pairs is small in the high temperature QGP. The interactions of heavy quarks are sensitive to medium dynamics. They decouple early in the evolution of QGP due to their large masses, thereby preserving the information from the system at early stage~\cite{nsm_roli,heavy_1,heavy_2,heavy_3,heavy_4,heavy_5,heavy_6,heavy_7}.
While traversing the hot and dense matter produced in nucleus--nucleus collisions, hard partons (partons with high transverse momentum $p_{T}$ ) produced in the early stages of the collision lose energy dominantly due to multiple scatterings and radiative energy loss. Hence, become quenched. Theoretical models predict that the mechanism as well as average energy loss will be different for heavy quarks compared to light quarks~\cite{mg,rb,mht}. Therefore, high $p_{T}$ charmed mesons ($D^{0}$, $D^{\pm}$, $D^{*}$, $D^{\pm}_{S}$ etc.) will show different suppression with respect to light mesons ($\pi$, $K$, $K_{S}^{0}$, etc.). In contrast, measurements of heavy flavor decay electrons at RHIC and charm hadrons at the LHC have shown significant suppression at high transverse momentum, $p_{T}$, similar to that of light hadrons for central collisions. Therefore, a complete understanding of the energy loss mechanisms in the QGP medium requires a systematic and precise measurements of the properties of various hadrons carrying different quark flavors at RHIC and the LHC. The dependence of the partonic energy loss on the in-medium path length is expected to be different for different energy loss mechanism. It is suggested that low-momentum heavy quarks could undergo hadronization both via fragmentation in the vacuum and recombination with other quarks from the medium~\cite{aa}. Azimuthal anisotropy measurements of the production of heavy--flavor hadron with respect to the reaction plane can be very useful in addressing these questions.

Elliptic flow ($v_{2}$) measured in heavy-ion collisions is believed to arise due to the pressure gradient developed when two nuclei collide at nonzero impact parameters followed by subsequent interactions among the constituents~\cite{hydro,hydro1,hydro2,hydro3,hydro4,early_v2}. The elliptic flow parameter is defined as the $2^{\mathrm {nd}}$ Fourier coefficient, $v_{2}$, of the particle distributions in emission azimuthal angle ($\phi$) with respect to the reaction plane angle ($\Psi$)~\cite{voloshinzhang,art,art2,v2_1,v2_2,v2_3,v2_4}
\begin{equation} 
\frac{dN}{d\phi} \propto 1+2 v_2\cos(2(\phi - \Psi)). 
\end{equation}
For a given rapidity window the second coefficient is given by
\begin{equation}
v_{2}=\langle\cos(2(\phi-\Psi))\rangle=\langle\frac{p_x^2-p_y^2}{p_x^2+p_y^2}\rangle,
\end{equation}
where $p_x$ and $p_y$ are the $x$ and $y$ components of the particle momenta. At small transverse momentum, $p_{T}$, a large $v_{2}$ is considered to be an evidence for the collective hydrodynamical expansion of the medium. Positive $v_{2}$, if observed at very high $p_{T}$ is expected to be due to path--length dependent energy loss by hard partons. Unlike light quarks and gluons, which can be produced or annihilated during the entire evolution of the medium, heavy quarks are expected to be produced mainly in initial hard scattering processes and their annihilation rate is small. Therefore, for all $p_{T}$, the final state heavy-flavor hadrons originate from heavy quarks that have experienced each stage of the system evolution.
The paper is organized in the following way. Section II describes sensitivity of heavy flavor hadron as probe of QCD medium. In Sec. III, elliptic flow of heavy-flavor decay electron are briefly discussed. Section IV and V describe elliptic flow of open charmed meson and  $J/\psi$, respectively, measured at RHIC and LHC. Comparisons between model and data are also presented in Sec. IV and V. Finally, we summarize in Sec. VI .

\section{Elliptic flow of heavy flavor as a sensitive probe}
\begin{figure}
\begin{center}
\includegraphics[scale=0.5]{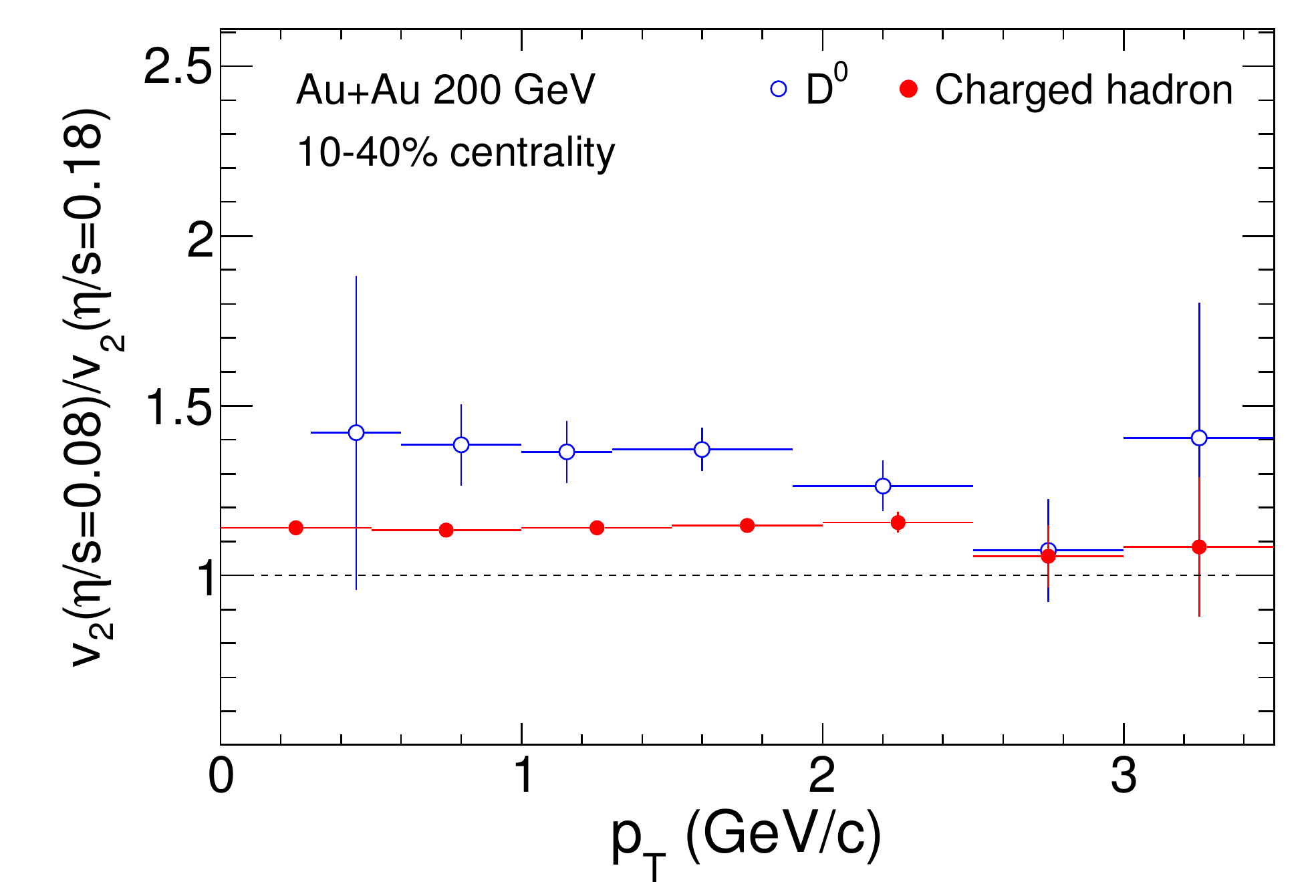}
\caption{(Color online) Ratio between $v_{2}$ for $\eta_s/s$ = 0.08  and for $\eta_s/s$ = 0.18  as a function of $p_{T}$ in Au+Au collisions at $\sqrt{s_{NN}}$ = 200 GeV for charged hadrons and $D^{0}$ for 10--40\% central collisions.}
\label{ratio_visco}
\end{center}
\end{figure}

We have recently studied the  elliptic flow of open charm mesons, using quark coalescence as a mechanism of hadronization within the framework of a multi-phase transport model (AMPT)~\cite{nsm_roli}. This study includes effect of partonic interaction cross-section, QCD coupling constant and specific viscosity on elliptic flow of open charm mesons within the transport model approach. The AMPT model is a hybrid transport model~\cite{ampt}. It uses the same initial conditions as in HIJING. In the AMPT model, the value of parton parton scattering cross-section, $\sigma_{PP}$, is calculated by 
\be
\sigma_{PP} \approx \frac{9 \pi \alpha_s ^2}{2 \mu^2},
\label{muaplha}
\ee
where  $\alpha_{s}$ and $\mu$ are the QCD coupling constant and screening mass respectively. 
Using the framework of AMPT model one can study the effect of specific viscosity on elliptic flow of hadrons. For a system of massless quarks and gluons at temperature  $T$ ($T = 378$ MeV at RHIC energy in AMPT~\cite{lhc_chgv2}), the specific viscosity is given by~\cite{lhc_chgv2}
\be
\frac{\eta_s}{s} \approx \frac{3\pi}{40 \alpha_s ^2} \frac{1}{\left( 9+ \frac{\mu^2}{T^2} \right) \ln \left( \frac{18 + \mu^2/T^2}{\mu^2/T^2} \right) - 18}. 
\label{eq_visco}
\ee
Hadronization of heavy quarks are not implemented in AMPT model. It only gives phase space information heavy quarks at freeze-out. We have implemented quark coalescence mechanism to form open charm mesons using phase space information of  quarks available from AMPT model. 
Within the framework of coalescence mechanism~\cite{coal}, the probability of producing a hadron from a soup of partons is determined by the overlap of the phase space distribution of partons at freeze-out with the parton Wigner phase space function inside the hadron.
The Wigner phase space function for quarks inside a meson is obtained from its constituent quark wave function~\cite{cmco}
\begin{small}
\bea
\rho^W(\mathbf{r},\mathbf{k}) & = & \int \psi \left( \mathbf{r}+\frac{\mathbf{R}}{2} \right) \psi^{\star}\left( \mathbf{r}-\frac{\mathbf{R}}{2} \right) \exp(-i \mathbf{k} \cdot \mathbf{R}) d^3 \mathbf{R} \nonumber \\
 & = & 8 \exp(-\frac{r^2}{\sigma^2}-\sigma^2 k^2)
\eea
\end{small}
where the relative momentum between the two quarks is $\mathbf{k} = (\mathbf{k}_1 - \mathbf{k}_2)/2$ and the quark wave function is given by spherical harmonic oscillator described as
\be
\psi (\mathbf{r}_1,\mathbf{r}_2) = \frac{1}{(\pi \sigma^2)^{3/4}} \exp \left[ \frac{-r^2}{2\sigma^2} \right]
\ee
with $\mathbf{r} = (\mathbf{r}_1 - \mathbf{r}_2)$ being the relative coordinate and $\sigma$ is the size parameter related to the root mean square radius as $\langle r^2 \rangle = (3/8)^{1/2} \sigma$. We have taken $\sigma = 0.47$ fm$^{2}$ from Ref.~\cite{cmco}.\\
We have used two different value of $\eta_s/s$ e.g. 0.08  and 0.18 by tuning input parameters of AMPT model  keeping parton-parton interaction cross-section equal to 10 mb. The values of $\alpha_s$ and $\mu$ for different value of $\eta_s/s$  are shown in table~\ref{tab1}.
\begin{table}
\caption{Values of $\eta_s/s$ for different values of $\alpha_s$ and $\mu$, keeping $\sigma_{PP}$ = 10 mb for Au+Au collisions at $\sqrt{s_{NN}}$ = 200 GeV.}
\begin{tabular}{|c|c|c|}
 \hline
 Specific viscosity, & QCD coupling constant,  &  Screening mass,   \\
 $\eta_s/s$ & $\alpha_s$ & $\mu$ (in fm$^{-1}$) \\ \hline
  0.08  &  0.47  &  1.77  \\ \hline
  0.18  &  0.23  &  0.88  \\ \hline
\end{tabular}
\label{tab1}
\end{table}
Fig.~\ref{ratio_visco} shows ratio between $v_{2}$ for $\eta_s/s$ = 0.08  and for $\eta_s/s$ = 0.18  as function of $p_{T}$ for 10-40$\%$ central Au+Au collisions at $\sqrt{s_{NN}}$ = 200 GeV. The red solid and open blue circles represents results for charged hadrons and $D^{0}$ respectively. We can see that $v_2$ decreases with increase in specific viscosity for both $D^{0}$ and  charged hadrons . This is consistent with the interpretation that increased sheer viscosity reduces transverse expansion due to increased interactions and hence, reduces $v_2$. We can also see that the change in $v_{2}$ for charged hadrons is $\sim$ 15$\%$, whereas for $D^{0}$ it lies between 30 -- 40$\%$ for $p_{T}$ $<$ 2.0 GeV/c. Therefore, the elliptic flow of open charmed meson is more sensitive to viscous properties of the QGP medium compared to light hadrons.  Comparison between our calculation and data has been discussed in Section IV.

\section{Elliptic flow of heavy-flavor decay electron}
\begin{figure}
\includegraphics[scale=0.40]{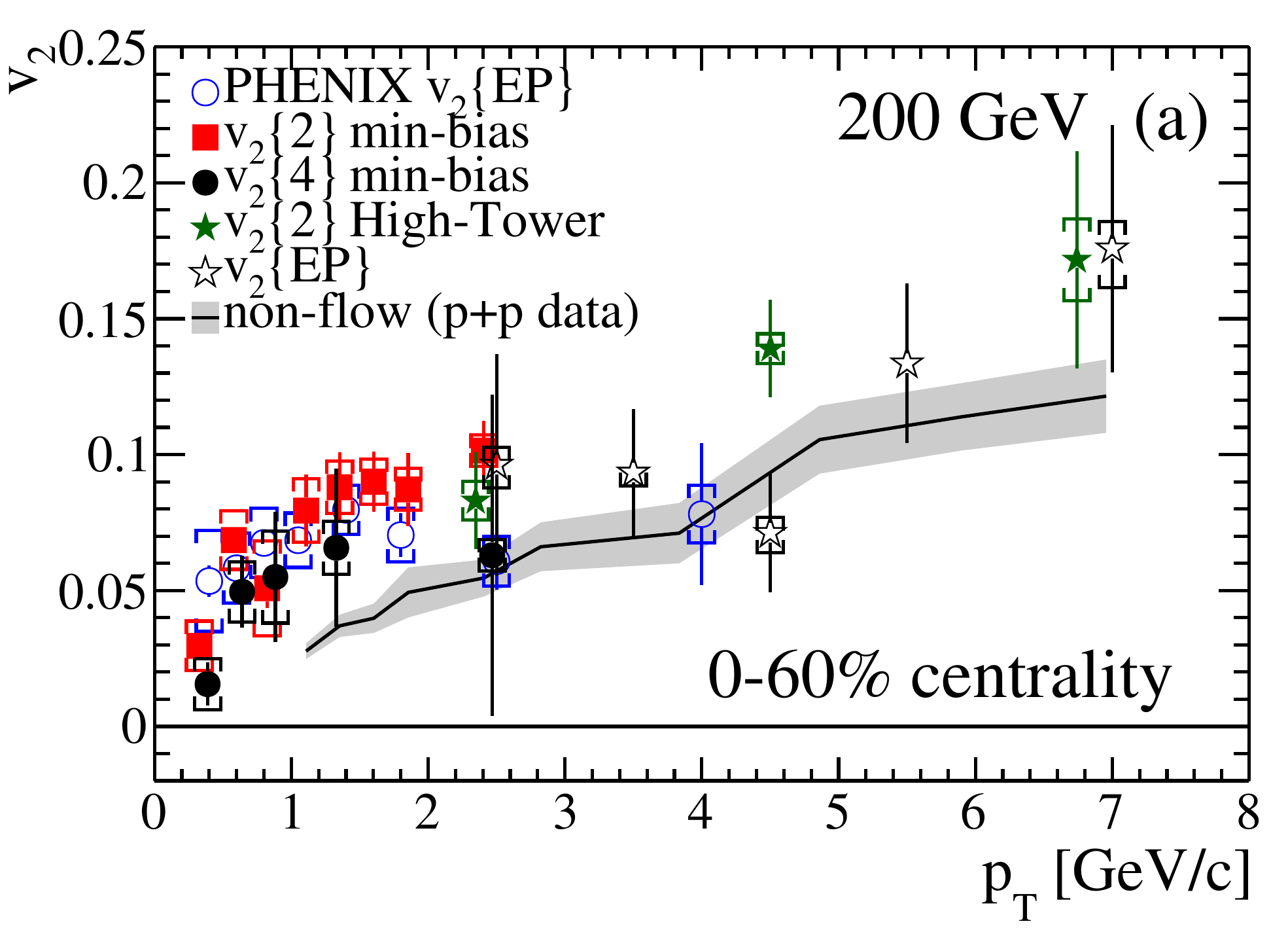}
\includegraphics[scale=0.40]{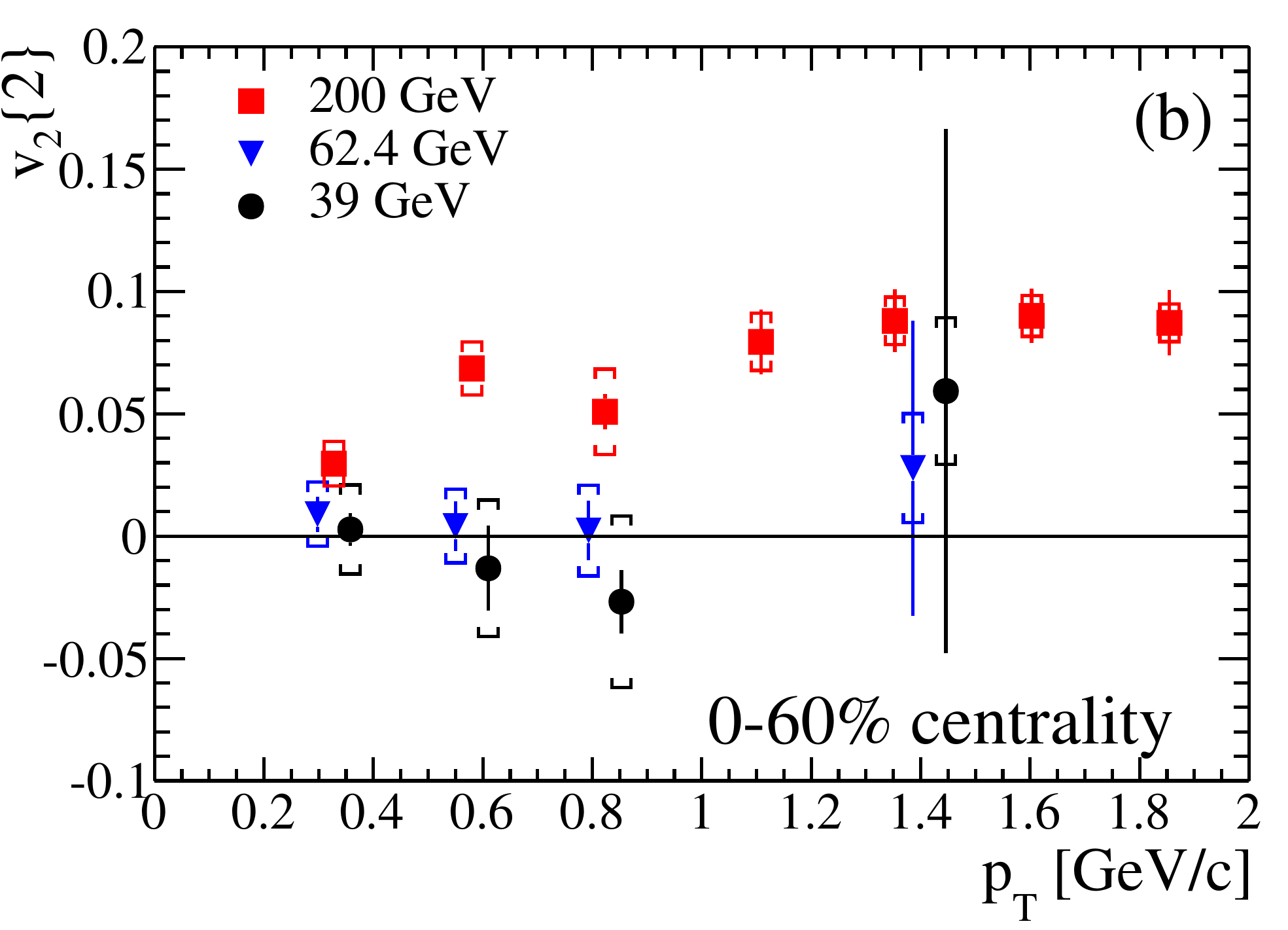}
\caption{ (Color online) (a) Comparison of azimuthal anisotropy of NPE at $\sqrt{s_{NN}}$ = 200 GeV measured by PHENIX~\cite{phenix_npe_prc} and STAR~\cite{star_npe_prc}. (b) NPE $v_{2}$ using two particle cumulant method at 200 and 62.4 and 39 GeV. The error bars represent the statistical uncertainty, and the brackets represent the systematic uncertainties.}
\label{star_npe}
\end{figure}

Measurement of electrons from semi-leptonic decay of heavy-flavor hadrons (also called non-photonic electrons, NPE)  is widely used to study heavy-flavor production in high-energy collisions. These NPE give the direction of the mother $D$ ($B$) mesons, especially when electron $p_{T}$ $>$ 1.5 (3) GeV/c. Thus, $v_{2}$ of NPE  serves as a proxy for $v_{2}$ of heavy quarks. Systematic measurements of the nuclear modification factor
($R_{\rm{AA}}$ and $R_{\rm{pA}}$) and the elliptic flow coefficient ($v_{2}$) of heavy-flavor decay electrons were performed at RHIC and LHC energies. Fig.~\ref{star_npe} shows the azimuthal anisotropy of NPE as a function of $p_{T}$ at $\sqrt{s_{NN}}$ = 39, 62.4 and 200 GeV as measured by STAR and PHENIX experiments~\cite{star_npe_prc,phenix_npe_prc}. A comparison between different methods for measurement of $v_{2}$ has been shown. The different methods show different sensitivity to non-flow and flow fluctuation. Non-zero positive $v_{2}$ has been observed for all methods. The increase of $v_{2}$ with $p_{T}$ for $p_{T}$ $>$ 4 GeV/c could be due an effect of jet-like correlations as the non-flow correlation, which is estimated from p+p collision, is of the similar order with measured $v_{2}$ in Au+Au collision at high $p_{T}$.  At 39 and 62.4 GeV, $v_{2}$ is consistent with zero as shown in Fig.~\ref{star_npe}(b). A very high precision measurement is required at 39 and 62.4 GeV, to understand NPE $v_{2}$ at these energies.

\begin{figure}
\begin{center}
\includegraphics[scale=0.40]{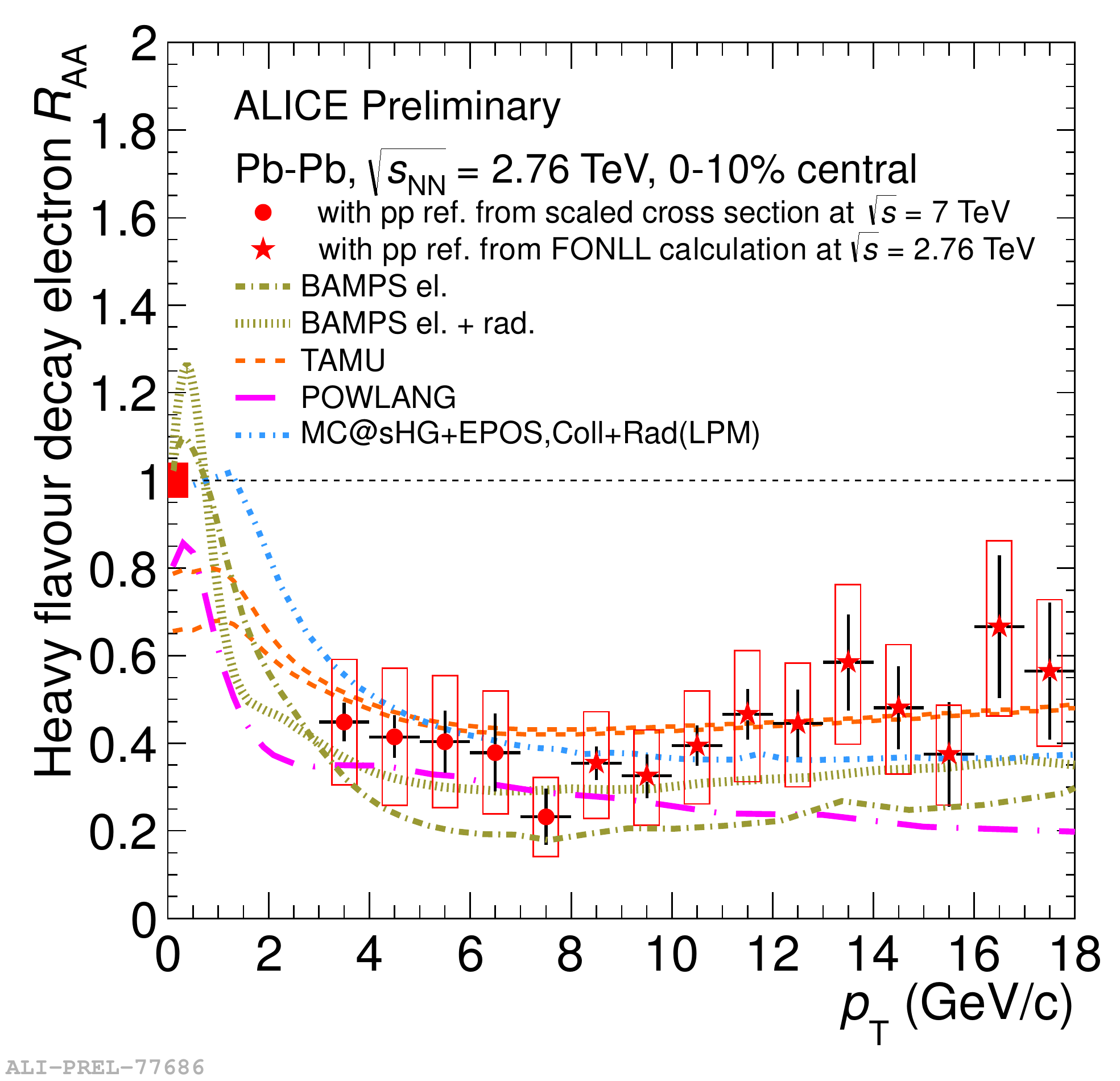}
\includegraphics[scale=0.40]{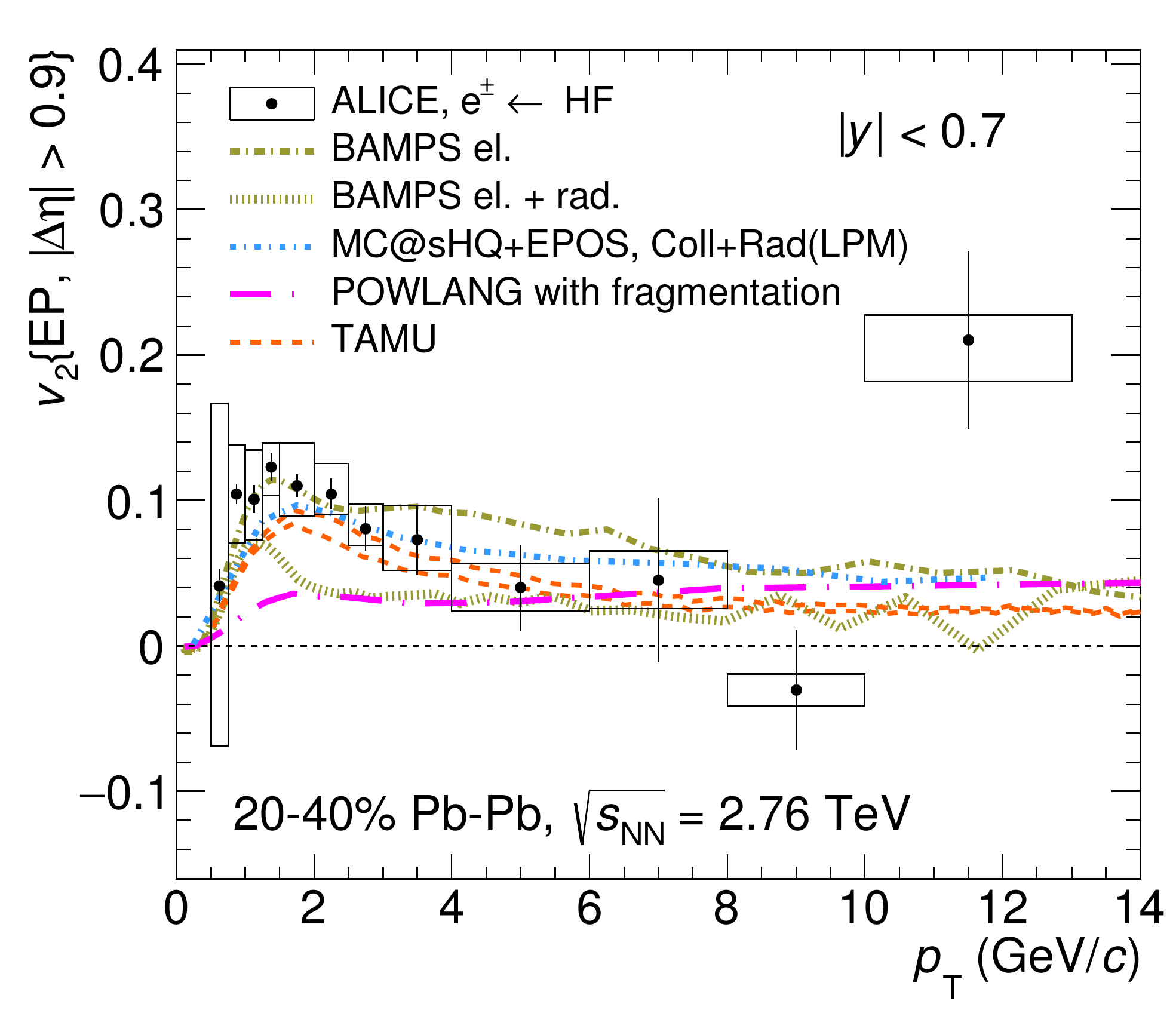}
\caption{ (Color online) Measured $v_{2}$ and $R_{AA}$ of NPE in Pb+Pb collision at 2.76 TeV~\cite{lhc_npe,lhc_npe1,lhc_npe2}. Theoretical prediction for $v_{2}$ and $R_{AA}$ of NPE are shown by lines~\cite{ju,rapp,mm,epos}. }
\label{alice_npe}
\end{center}
\end{figure}

The nuclear modification factors ($R_{AA}$) and elliptic flow of NPE  in Pb+Pb collisions at 2.76 TeV is shown in Fig.~\ref{alice_npe}~\cite{lhc_npe}. A finite positive $v_{2}$ of NPE is observed for $p_{T}$ $<$ 6 GeV/c in Pb+Pb collision at 2.76 TeV, quite similar to Au+Au collision at 200 GeV at RHIC. Large positive $v_{2}$ of NPE at low $p_{T}$  might indicate that charm quarks participate in the collective expansion of the dense and hot QGP. Also, a strong suppression of yield of NPE is observed for $p_{T}$ $>$ 3 GeV/c in 0-10$\%$ most central Pb+Pb collisions. 

The results from the models calculations~\cite{ju,rapp,mm}, that include parton energy loss in the hot and dense QCD medium, are shown as lines for both $v_{2}$ and $R_{AA}$ in Fig.~\ref{alice_npe}. The simultaneous description of the measured $v_{2}$ and $R_{AA}$ is challenging for models. BAMPS~\cite{ju} gives a good description of NPE $v_{2}$ but predicts a larger in-medium suppression than measured. In BAMPS approach, heavy quarks are transported through the medium while undergoing collisional and radiative energy loss. The prediction from POWLANG~\cite{mm} describes the NPE $R_{AA}$  but their calculation underestimates NPE $v_{2}$. In POWLANG, heavy quarks are transported following a Langevin approach and consider collisional energy loss only. The prediction from Rapp et al~\cite{rapp} (TAMU) and MC@sHQ+EPOS, Coll+Rad(LPM)~\cite{epos}  describes the NPE $R_{AA}$ and  $v_{2}$ reasonably well.  Model calculation by Rapp et al includes in-medium resonance scattering and coalescence of heavy quarks in the medium. The MC@sHG+EPOS model includes radiative and collisional energy loss in an expanding medium based on the EPOS model.

\section{Elliptic flow of open charmed meson}
\subsection{Available Experimental Data}
\begin{figure}
\begin{center}
\includegraphics[scale=0.80]{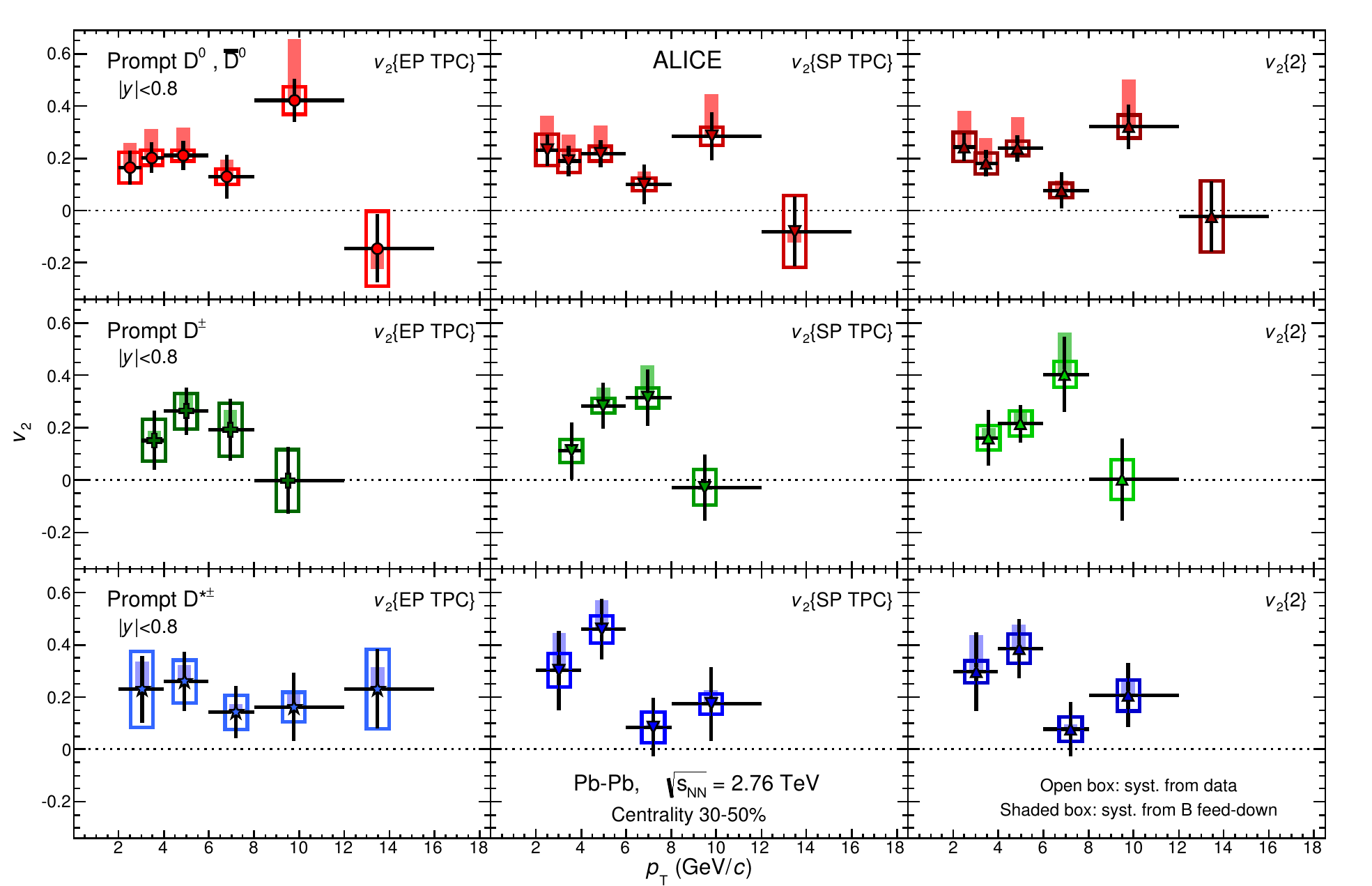}
\caption{ (Color online)  Elliptic flow as a function of $p_{T}$ in the 30-50$\%$ centrality bin, for $D^{0}$, $D^{+}$ and $D^{*+}$ mesons with the event plane, scalar product and two-particle cumulant methods~\cite{alice_d0_prc}. The vertical error bars represent the statistical uncertainty, the open boxes are the systematic uncertainties.}
\label{alice_d0dstardp}
\end{center}
\end{figure}

The elliptic flow of $D$ mesons at mid-rapidity ($|y|$ $<$ 0.8) has been measured $p_{T}$ differentially in Pb-Pb collisions by the ALICE at LHC~\cite{alice_d0_prc}. The transverse momentum dependence of the $v_{2}$ of $D^{0}$, $D^{+}$ and $D^{*+}$ mesons in the 30-50$\%$ collisions centrality at $\sqrt{s_{NN}}$ = 2.76 TeV are shown in Fig.~\ref{alice_d0dstardp}. Measurements were done using three different methods, namely event plane, scalar product and two-particle cumulant methods. Results from event plane method is shown in the left column of Fig.~\ref{alice_d0dstardp}. The central and right panels of the Fig.~\ref{alice_d0dstardp} show the $v_{2}$ results obtained with the scalar product and two-particle cumulant methods respectively. The event plane is estimated from TPC tracks within $|\eta|$ $<$ 0.8.  For the other methods, TPC tracks in $|\eta|$ $<$ 0.8 were used as reference particle. The elliptic flow of $D^{0}$, $D^{+}$ and $D^{*+}$ mesons are consistent within uncertainties. At very high $p_{T}$  ($p_{T}$ $>$ 12 GeV/c), $v_{2}$ is consistent with zero within the large statistical uncertainties. In the $p_{T}$ range between 2 $<$ $p_{T}$ $<$ 6 GeV/c, the measured $v_{2}$ is found to be larger than zero with 5.7 $\sigma$ significance. It suggests that low $p_{T}$ charm quarks possibly participate in the collective expansion of the medium. However, the possibility that the observed $D$-meson $v_{2}$ is completely due to the contribution from light-quark in a scenario with hadronization via recombination cannot be ruled out. We need high precision data at low $p_{T}$ and more theoretical understanding about the charm quark hadronization to understand the origin of collectivity of measured $D$-mesons $v_{2}$.\\
  
\begin{figure}
\begin{center}
\includegraphics[scale=0.80]{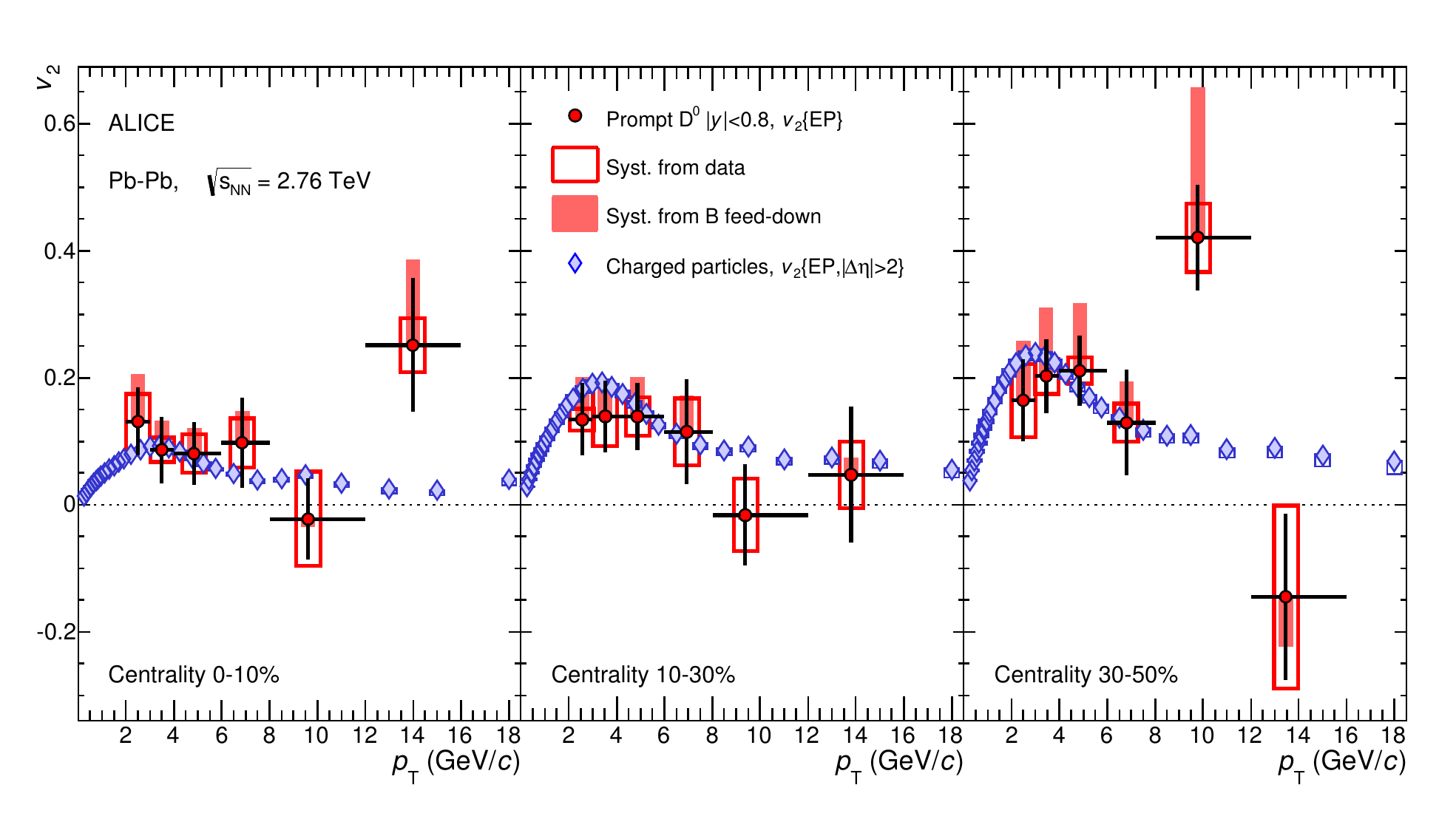}
\caption{ (Color online)  Comparison of $D^{0}$ meson~\cite{alice_d0_prc} and charged-particle~\cite{alice_chg_prl} $v_{2}$ in three centrality classes as a function of $p_{T}$. The vertical error bars represent the statistical uncertainty, the open boxes are the systematic uncertainties.}
\label{alice_d0chg}
\end{center}
\end{figure}

The $p_{T}$ dependence of $D^{0}$ $v_{2}$ in the three centrality classes 0-10$\%$, 10-30$\%$ and 30-50$\%$ are presented in Fig.~\ref{alice_d0chg}~\cite{alice_d0_prc}. $v_{2}$ of charged hadrons are also shown for comparison~\cite{alice_chg_prl}. Both the measurements are done with the event plane method. For these three centrality classes, the $D^{0}$ meson $v_{2}$ is comparable in magnitude to that of inclusive charged hadrons. These results indicate that the interactions with the medium constituents transfer information of the azimuthal anisotropy of the system to the charmed particles.

\begin{figure}
\includegraphics[scale=0.40]{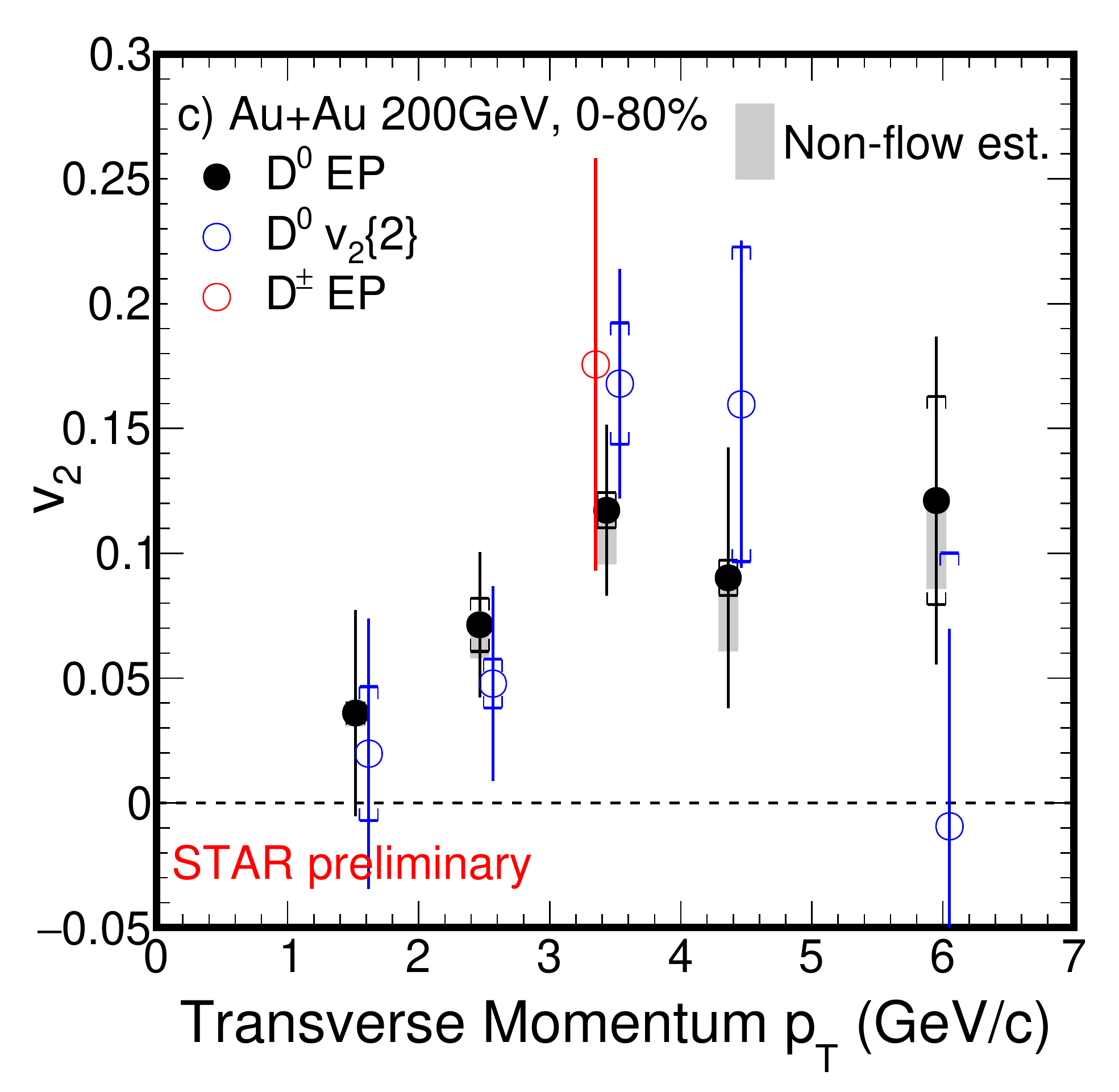}
\includegraphics[scale=0.40]{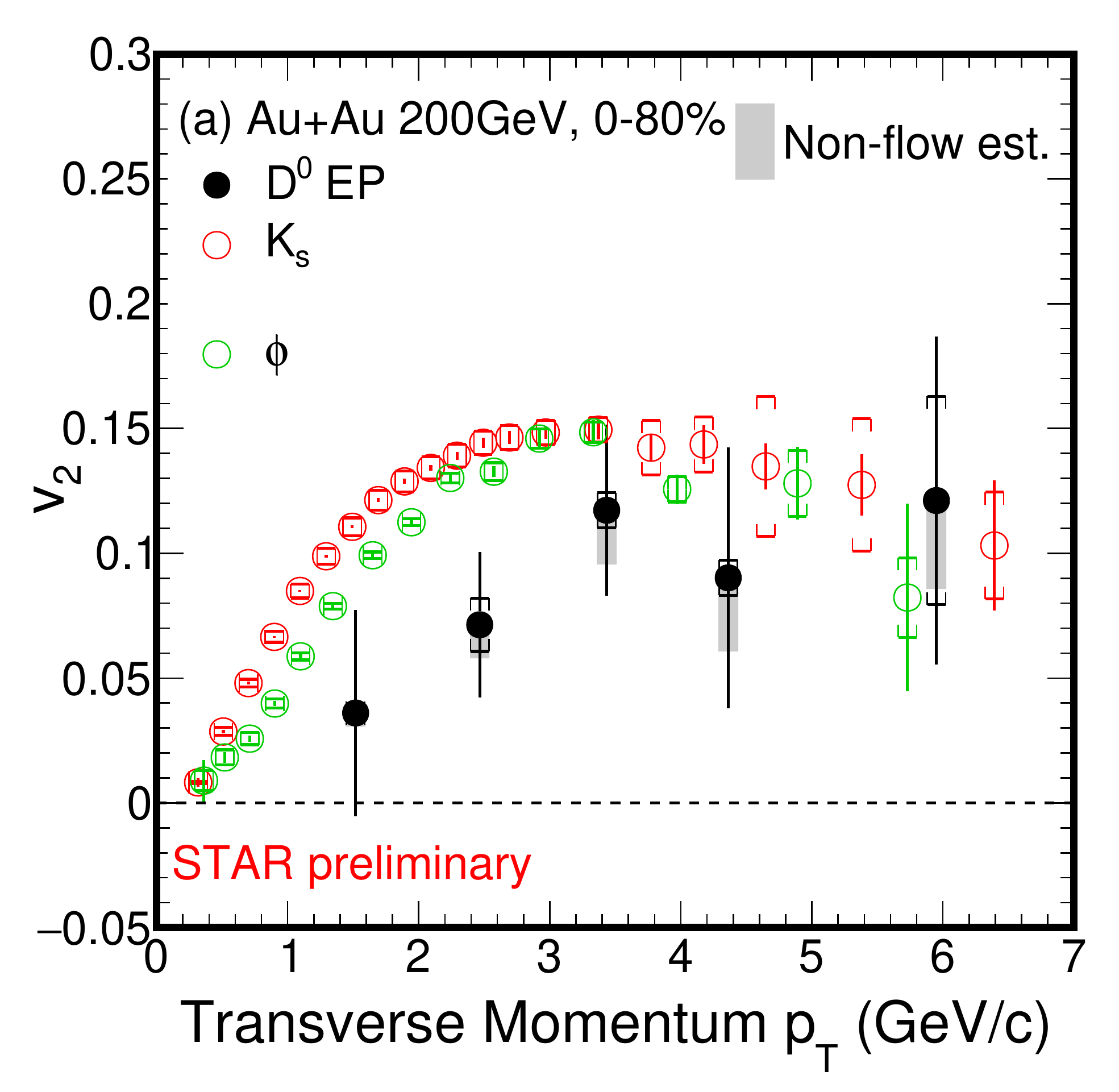}
\caption{ (Color online) Left panel: The elliptic flow of $D^{0}$ and $D^{+}$ meson as a function of $p_{T}$ in minimum bias (0--€"80$\%$) Au+Au collisions at $\sqrt{s_{NN}}$ = 200 GeV ~\cite{star_d0_v2_ml}. Right panel:  $v_{2}$ for $D^{0}$ compared to that of light mesons ($K^{0}_{S}$ and $\phi$) ~\cite{star_k0s_prc,star_phi_prl2016}. The vertical error bars represent the statistical uncertainty, the cap symbols are the systematic uncertainties.}
\label{star_d0dp}
\end{figure}

The STAR experiment at RHIC has also reported the first preliminary results of $D^{0}(p_{T})$ $v_{2}$ at mid-rapidity ($|y|$ $<$ 1.0) in Au+Au collision at $\sqrt{s_{NN}}$ = 200 GeV using newly installed Heavy Flavor Tracker (HFT)~\cite{star_d0_v2_ml,star_d0_v2_nsm}. The $D$-meson $v_{2}$ measured by STAR in minimum bias (0-80$\%$) Au+Au collisions at $\sqrt{s_{NN}}$ = 200 GeV is shown in left panel of
Fig.~\ref{star_d0dp}~\cite{star_d0_v2_ml}. Measurements are done at mid-rapidity ($|y|$ $<$1.0). The blue and black data points are the $D^{0}$ $v_{2}$ measured using two-particle correlation and event plane method respectively. Results from both the method are consistent within statistical uncertainty. The $D^{\pm}$ $v_{2}$ is shown by red symbol and calculated using event plane method. $D^{0}$ azimuthal anisotropy is non-zero for  2 $<$ $p_{T}$ $<$ 5 GeV/c. The right panel of Fig.~\ref{star_d0dp} shows the comparison of $D^{0}$ $v_{2}$ with other mesons species ($K^{0}_{S}$ and $\phi$). It seems that $D^{0}$ $v_{2}$ for $p_{T}$ $<$ 4 GeV/c is systematically lower than $K^{0}_{S}$~\cite{star_k0s_prc} and $\phi$~\cite{star_phi_prl2016}, but one should be very careful while comparing different particle species for a wide centrality bin e.g. 0-80$\%$ centrality bin as the production of heavy open charmed meson are more biased towards central collisions than light hadrons like $K^{0}_{S}$ and $\phi$~\cite{nsm_roli}. 

\subsection{Model Comparisons at LHC Energy}
\begin{figure}
\includegraphics[scale=0.40]{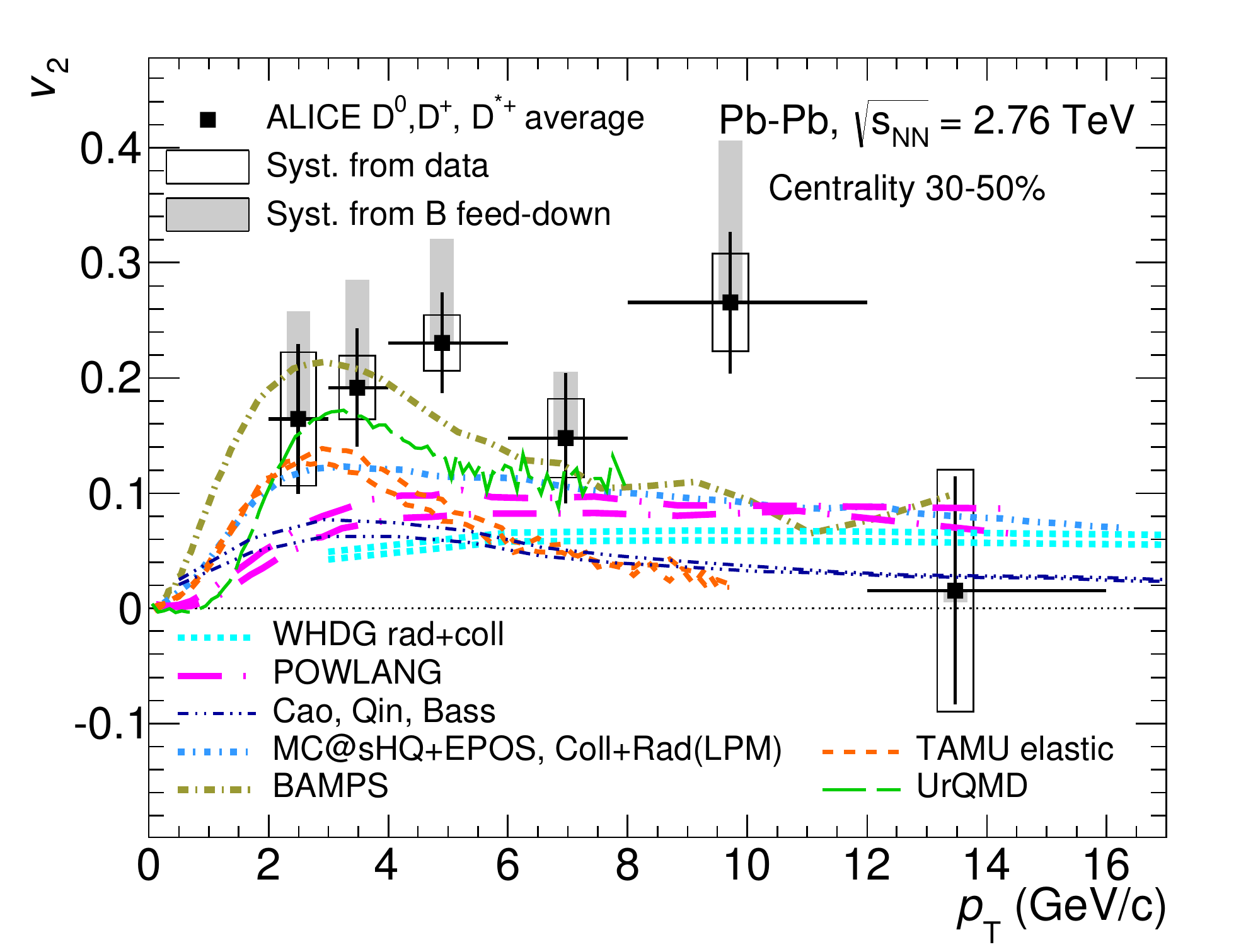}
\includegraphics[scale=0.40]{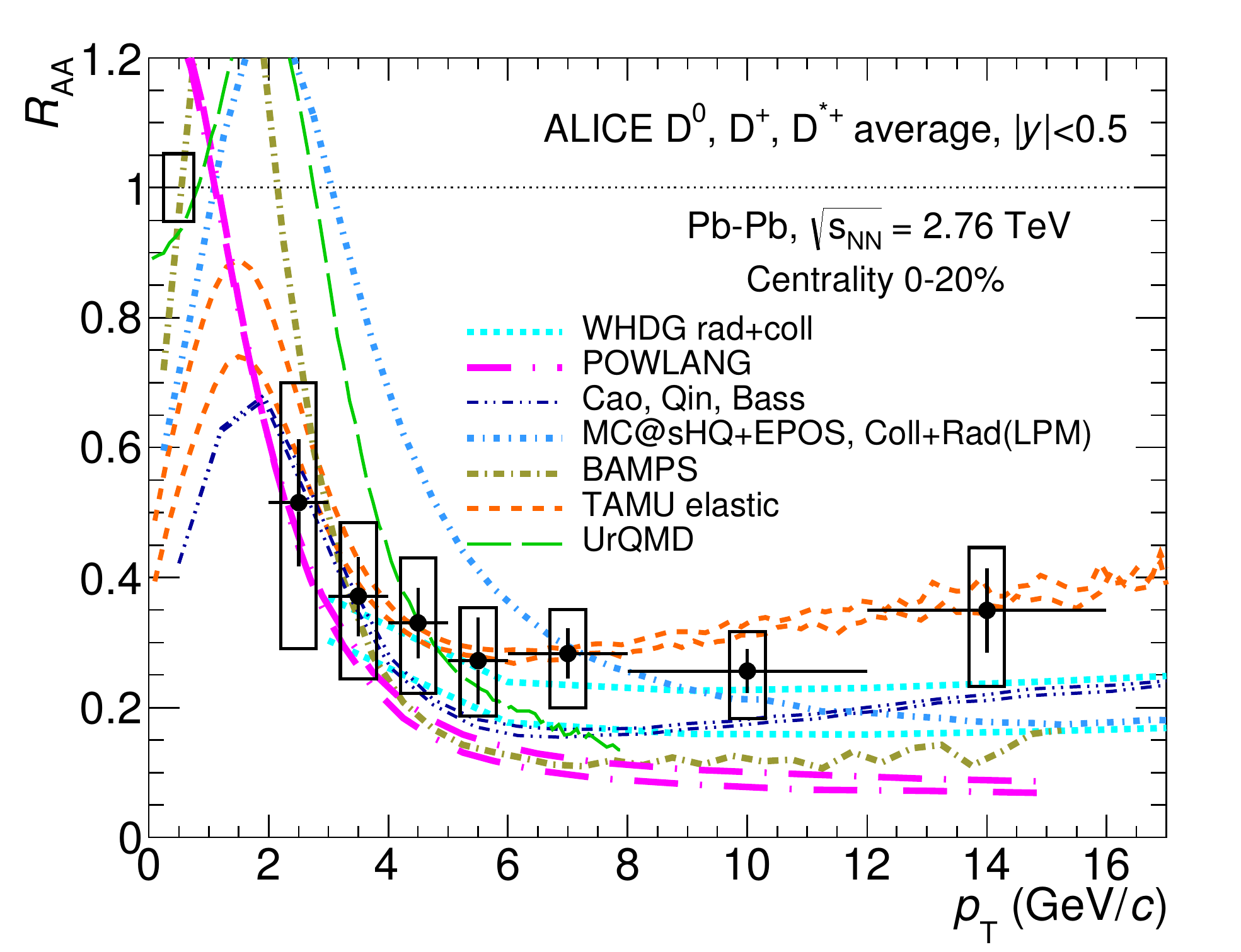}
\caption{ (Color online) $D$-meson $v_{2}$ and $R_{AA}$ in 30-50$\%$  semi-central Pb-Pb collisions at $\sqrt{s_{NN}}$  = 2.76 TeV and comparison with selected theoretical models~\cite{whdg_model,cao_model,bapms,powlong,mchq,tamu,urqmd}. }
\label{alice_davg_v2_raa_model}
\end{figure}

Various observables are compared to theoretical calculations to understand the physical mechanism behind the measurements. In this section, we will discuss the most recent theoretical baseline calculations for the elliptic flow of open charm meson.  Simultaneous description of the measured $v_{2}$ and $R_{AA}$ by theoretical model is a challenging job and is an open issue~\cite{skd1,skd2,whdg_model,cao_model,bapms,powlong,mchq,tamu,urqmd}. In Fig.~\ref{alice_davg_v2_raa_model} the $v_{2}$ (in 30-50$\%$ central collisions) and $R_{AA}$ (in 0-20$\%$ central collisions) of D mesons (average of $D^{0}$, $D^{+}$, and $D^{*+}$) in Pb-Pb collisions at $\sqrt{s_{NN}}$ = 2.76 TeV is shown and compared to selected model predictions.

\subsubsection{Coalescence Based Models}
The model by Cao et al. is  based on the Langevin approach where the space-time evolution of the medium is modeled using viscous hydrodynamic. In this model, hadronization is done using quark coalescence mechanism. This model describes  $R_{AA}$ in central collisions very well, but tend to underestimate the $v_{2}$ at low $p_{T}$. The model~\cite{mchq}, labeled as MC@sHQ+EPOS, Coll+Rad(LPM), is a perturbative QCD (pQCD) model that includes collisional and radiative energy loss mechanisms for heavy quarks. Hadronization is performed via quark recombination in this model.  It underestimates the low-$p_{T}$ suppression, but yields a substantial anisotropy ($\sim$10$\%$) which slightly underestimates observed data. It correctly describes high-$p_{T}$ suppression. In the TAMU model~\cite{tamu}, heavy-quark transport coefficient is calculated within a non-perturbative T-matrix approach. This model includes hydrodynamic medium evolution and quark coalescence as a mechanism of hadronization.  This model provides a good description of the observed suppression of D mesons over the entire $p_{T}$ range. However, it fails to reproduces observed anisotropy for $p_{T}$ $>$ 4 GeV/c.
The Ultra relativistic Quantum Molecular Dynamics (UrQMD)~\cite{urqmd} model is based on a microscopic transport theory where the phase space description of the reactions are important. The hybrid UrQMD model includes a realistic description of the medium evolution by combining hadronic transport and ideal hydrodynamics. Hadroniztion via quark recombination is implemented.  The model describes the measured anisotropy and suppression in the interval 4 $<$ $p_{T}$ $<$ 8 GeV/c, but fails to explain the data for very low and high $p_{T}$ region.
\subsubsection{Fragmentation Based Models}
In WHDG model, the observed anisotropy results from path-length dependent energy loss and hadronization is performed using vacuum fragmentation function. This model describes $R_{AA}$ in central collisions reasonably well, but tend to underestimate the $v_{2}$ at low $p_{T}$.
BAMPS model is a partonic transport model which includes multi-parton scattering based on Boltzmann approach. Like WHDG model, in BAMPS, hadronization is performed using vacuum fragmentation functions. BAMPS model describes  both  $R_{AA}$ and $v_{2}$ reasonably well. 
POWLANG~\cite{powlong} is also a transport model which is based on collisional processes treated within the framework of Langevin dynamics. Hadronization, in this model, is done using vacuum fragmentation functions. This model overestimates the high-$p_{T}$ suppression, and significantly underestimates observed $v_{2}$ at low $p_{T}$. \\
In summary, models including hadronization of charm quarks from recombination with light quarks from the medium (e.g. TAMU) provide a better description of the data at low transverse momentum.  

\subsection{Model Comparisons at RHIC Energy}
\begin{figure}
\includegraphics[scale=0.40]{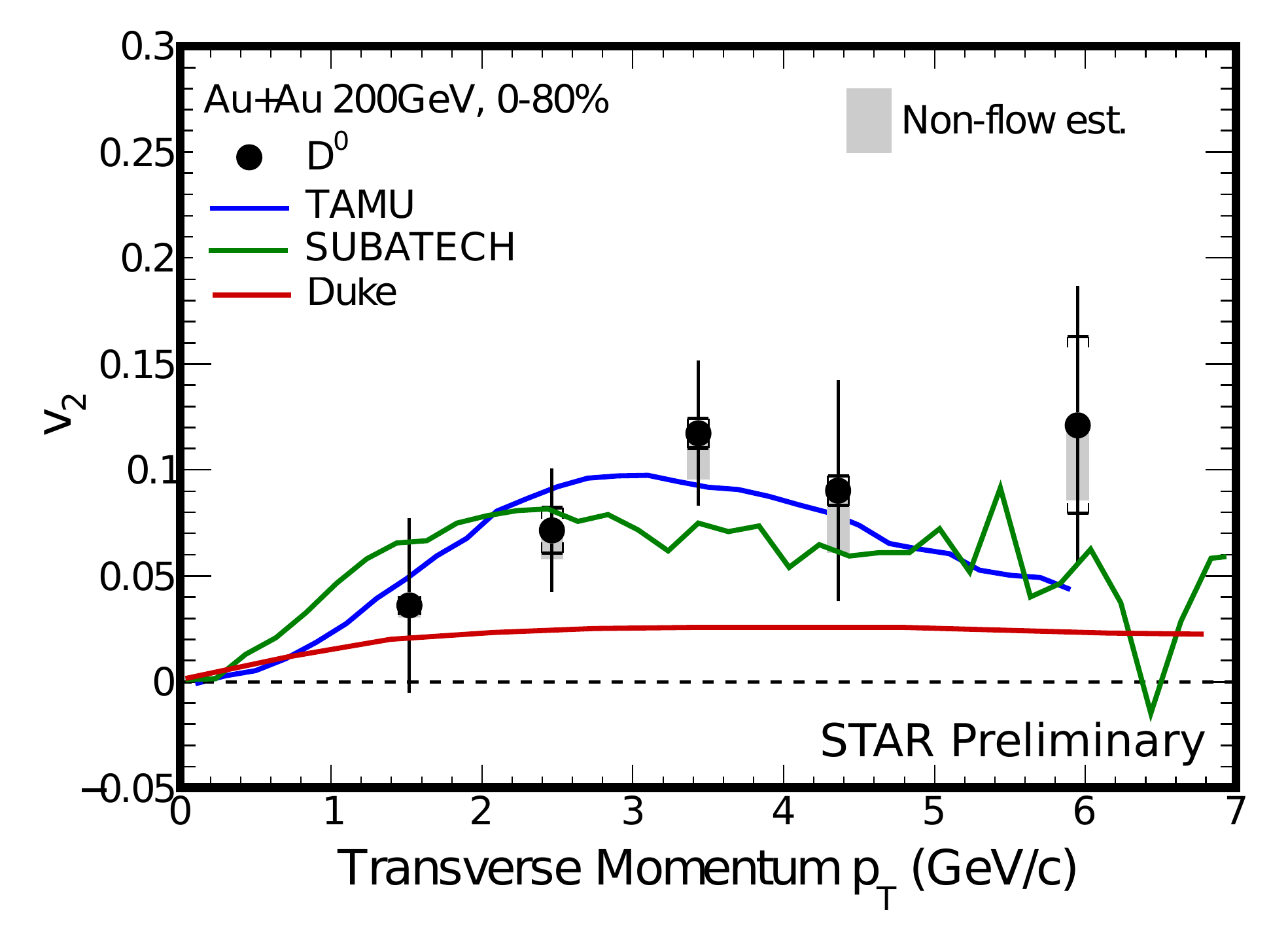}
\includegraphics[scale=0.40]{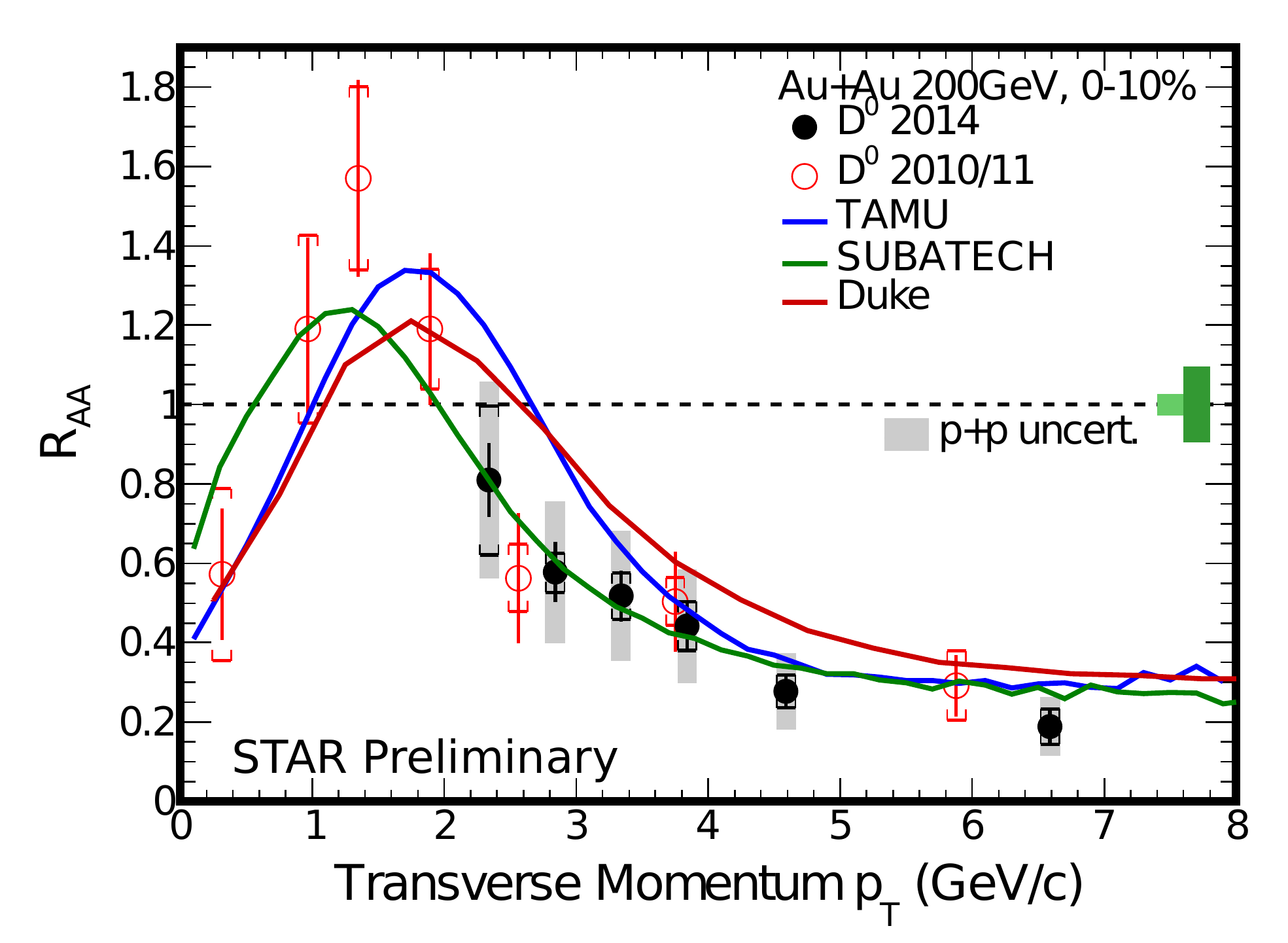}
\caption{ (Color online) $D^{0}$-meson $v_{2}$ (0-80$\%$)~\cite{star_d0_v2_ml,star_d0_v2_nsm} and $R_{AA}$ (0-10$\%$)~\cite{star_d0_v2_xie} in Au+Au collisions at $\sqrt{s_{NN}}$  = 200 GeV and comparison with selected theoretical models~\cite{subatech,tamu,cao_model}. }
\label{star_davg_v2_raa_model}
\end{figure}

Figure~\ref{star_davg_v2_raa_model} show $v_{2}$  and
$R_{AA}$ of $D^{0}$-meson for 0-80$\%$ and 0-10$\%$ centrality respectively in Au+Au collisions at $\sqrt{s_{NN}}$  = 200 GeV with selected theoretical model predictions. The physics of DUKE model by Cao et al. and TAMU is already discussed before. The SUBATECH~\cite{subatech} model based on pQCD calculation with the diffusion coefficient parameter $\sim$ 2-4. These three models can describe $D^{0}$ suppression reasonable well, however DUKE model underestimates measured anisotropy. The TAMU and SUBATECH describe  $D^{0}$-meson $v_{2}$ data reasonably well. 

\begin{figure}
\includegraphics[scale=0.4]{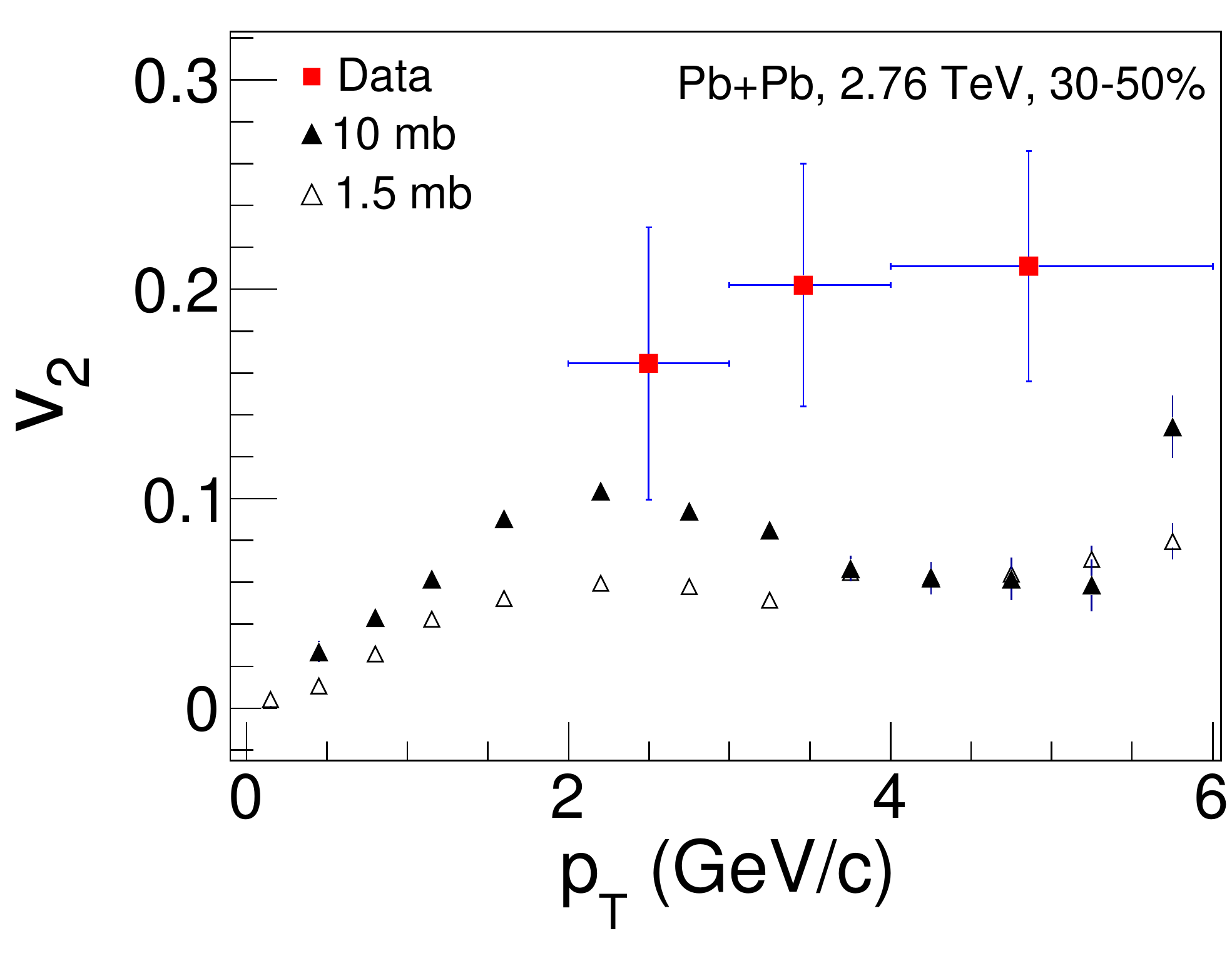} 
\includegraphics[scale=0.4]{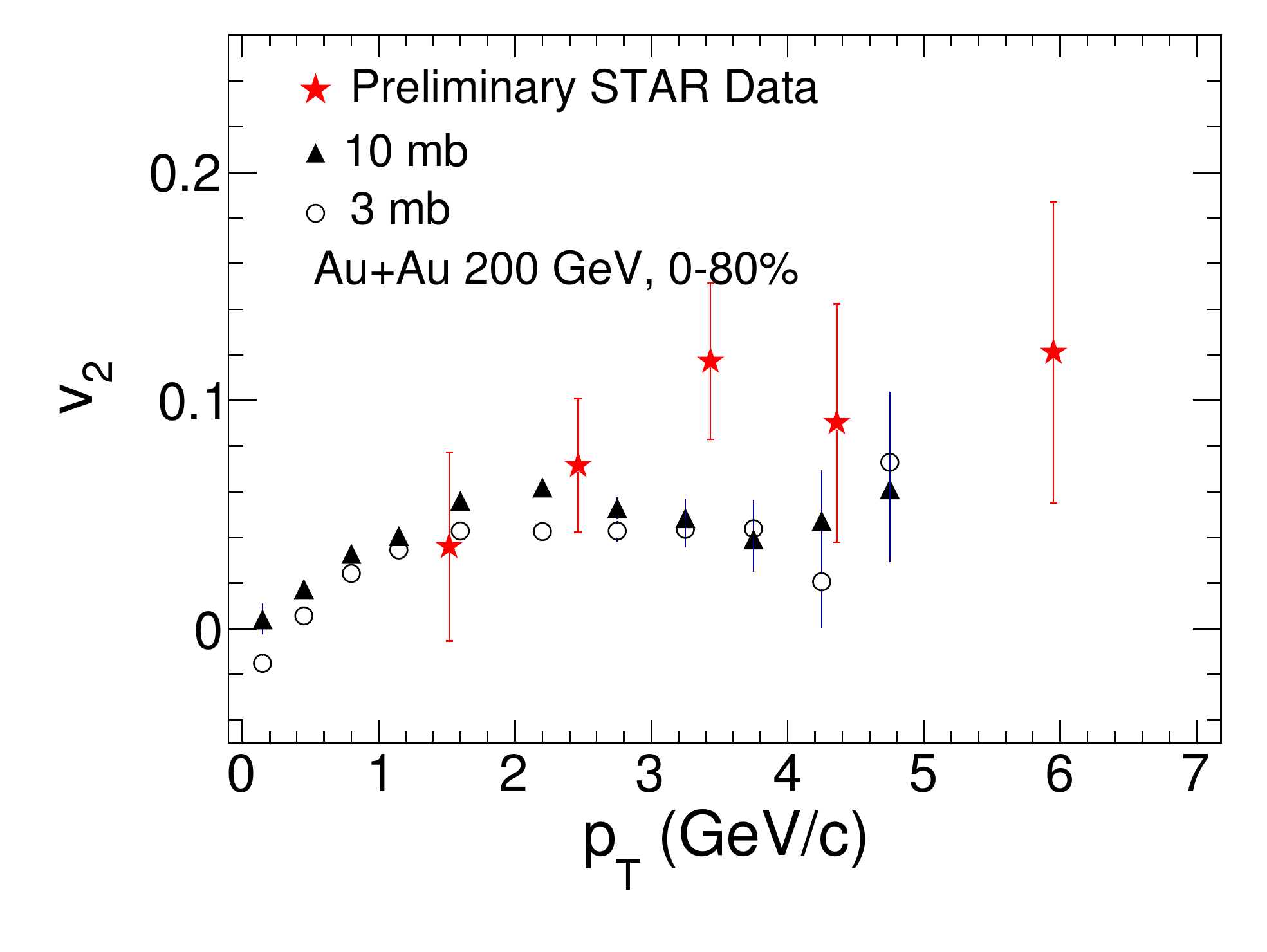}
\caption{(Color online) Left panel: Elliptic flow of  $D^{0}$ meson at mid-rapidity in Pb+Pb collision at $\sqrt{s_{NN}}$ = 2.76 TeV for 30-50$\%$ centrality and model predictions from AMPT. Only statistical error is shown for ALICE data. Right Panel: Elliptic flow of  $D^{0}$ meson at mid-rapidity in 0-80$\%$ min-bias Au+Au collision at $\sqrt{s_{NN}}$ = 200 GeV and prediction from AMPT. Only statistical error is shown for STAR preliminary data.}
\label{lhcv2_ampt}
\end{figure}

Fig.~\ref{lhcv2_ampt} (left panel) shows comparison between our AMPT model calculations~\cite{nsm_roli} for $D^{0}$-meson  $v_{2}$ and measured $D^{0}$-meson $v_{2}$ at 2.76 TeV for 30-50$\%$ central collisions by the ALICE experiment. Here  $\sigma_{PP}$ is taken to be 1.5 mb and 10mb with other parameters tuned for LHC data (charged hadron $v_{2}$ and multiplicity). Previous study shows that 1.5 mb parton parton scattering cross-section is sufficient to described charged hadron $v_{2}$ at mid-rapidity for $p_{T}$ $<$ 2.0 GeV/c. However, we find that cross-sections of both both 1.5 and 10 mb underestimate LHC $D^{0}$-meson $v_{2}$ data. It would be very interesting to see how the data and model behave at low $p_{T}$ (below 2 GeV/c). Therefore, the results from future ALICE upgrade~\cite{alice_upgrade} will be very useful to study both heavy flavor and charged hadrons $v_{2}$ at low $p_{T}$. The right panel of Fig.~\ref{lhcv2_ampt} shows the comparison between $D^{0}$ $v_{2}$ with AMPT model predictions at top RHIC energy. AMPT model calculation roughly explain data within large statistical errors. 

\section{Elliptic flow of $J/\psi$}
The $J/\psi$ meson is bound state of charm ($c$) and anti-charm ($\bar{c}$) quark. It was discovered independently by two research groups on 11 November 1974~\cite{jpsi_disco_1,jpsi_disco_2}. The importance of this discovery is highlighted by the fact that the subsequent rapid changes in high-energy physics around that time came to be collectively known as the "November Revolution". In relativistic heavy-ion collisions, $J/\psi$ can be produced mainly by recombination of charm ($c$) and anti-charm and/or direct pQCD processes. By measuring anisotropic flow of $J/\psi$, one may infer the relative contribution of $J/\psi$ particles from recombination and from direct pQCD processes. $J/\psi$ produced from quark recombination will inherit the flow of charm quarks. On the other-hand, if $J/\psi$ is produced from direct pQCD processes, it should have very little $v_{2}$. A detailed comparison between experimental measurements and models on $J/\psi$ $v_{2}$ will be helpful to understand the production mechanism of $J/\psi$.

\begin{figure}
\begin{center}
\includegraphics[scale=0.4]{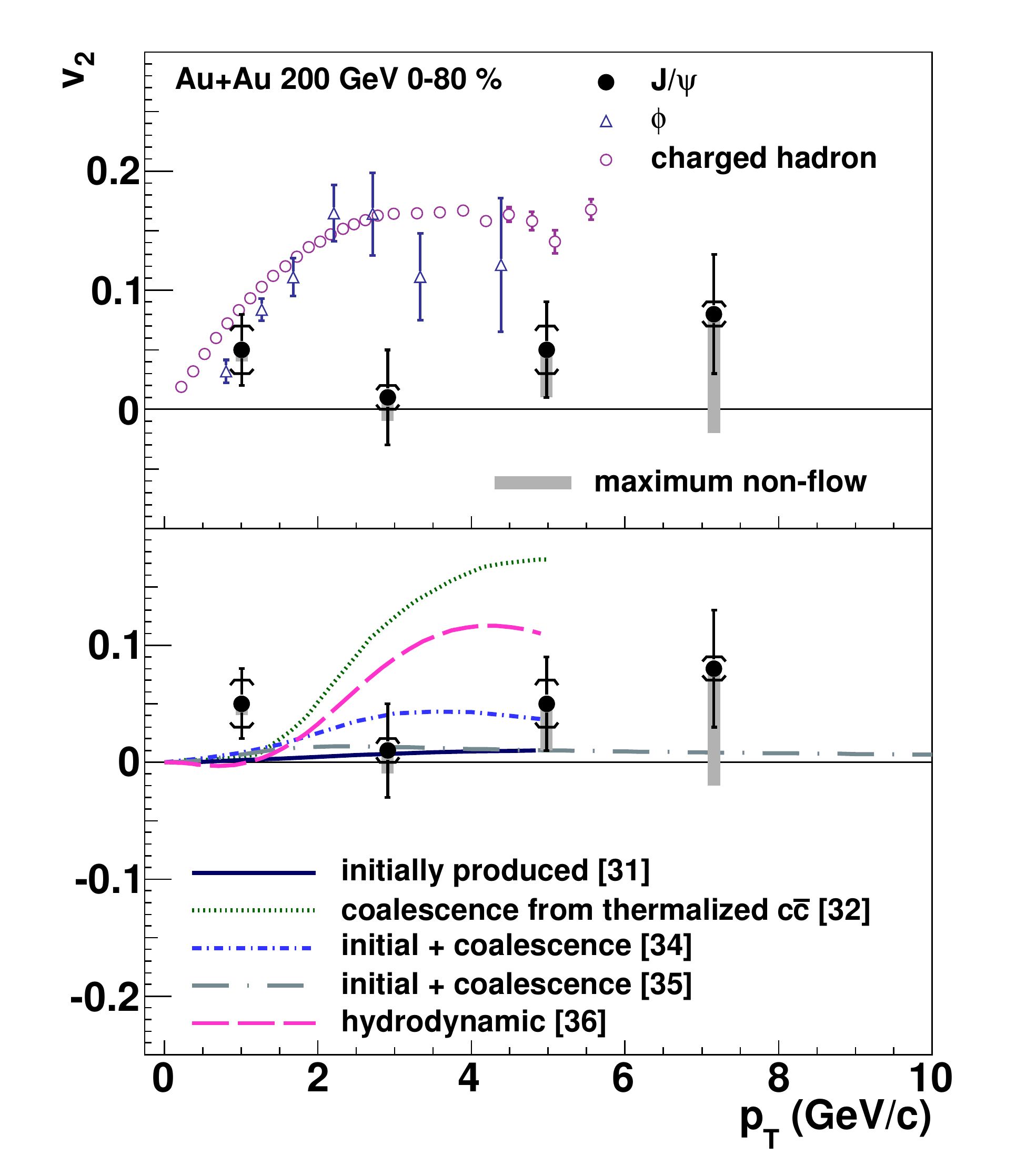}
\caption{(Color online) 
Elliptic flow of for $J/\psi$ particles at mid-rapidity  in 0-80$\%$ min-bias Au+Au events at $\sqrt{s_{NN}}$ =200 GeV~\cite{jpsi_starprl} compared with charged hadrons~\cite{chg_starprl} and the $\phi$ meson~\cite{phi_starprl} (upper panel) and theoretical calculations~\cite{jpsi_thermmodel,jpsi_pQCDmodel,jpsi_pqcd_thermmodel_1,jpsi_pqcd_thermmodel_2} (lower panel).}
\label{jpsiv2_star}
\end{center}
\end{figure}

Figure~\ref{jpsiv2_star} show elliptic flow of for $J/\psi$ at mid-rapidity in 0-80$\%$ min-bias Au+Au events at $\sqrt{s_{NN}}$ =200 GeV ~\cite{jpsi_starprl} compared with charged hadrons~\cite{chg_starprl} and $\phi$ meson~\cite{phi_starprl} in the upper panel and with theoretical calculations in the lower panel.  $J/\psi$ $v_{2}$ is found to be very small in comparison to that of charged hadrons and $\phi$-meson. Models which include $J/\psi$ production from coalescence of thermalized  $c\bar{c}$~\cite{jpsi_thermmodel}, gives the  maximum of $J/\psi$ $v_{2}$ to be almost the same in magnitude as light hadrons. $v_{2}$ of $J/\psi$ produced from initial pQCD processes~\cite{jpsi_pQCDmodel} is predicted to be very small compared to light hadrons. Models that include $J/\psi$ production from both initial pQCD process and coalescence mechanism~\cite{jpsi_pqcd_thermmodel_1,jpsi_pqcd_thermmodel_2} also give much smaller $J/\psi$ $v_{2}$ in comparison to light hadrons.  

In summary, models thats include $J/\psi$ production from both initial pQCD process and coalescence production or entirely from initial pQCD process describe the data better at top RHIC energy. At this point, it is still unclear and we would need very high precision measurements to estimate the fraction of the total $J/\psi$ yield that comes from pQCD and coalescence processes.

\begin{figure}
\begin{center}
\includegraphics[scale=0.4]{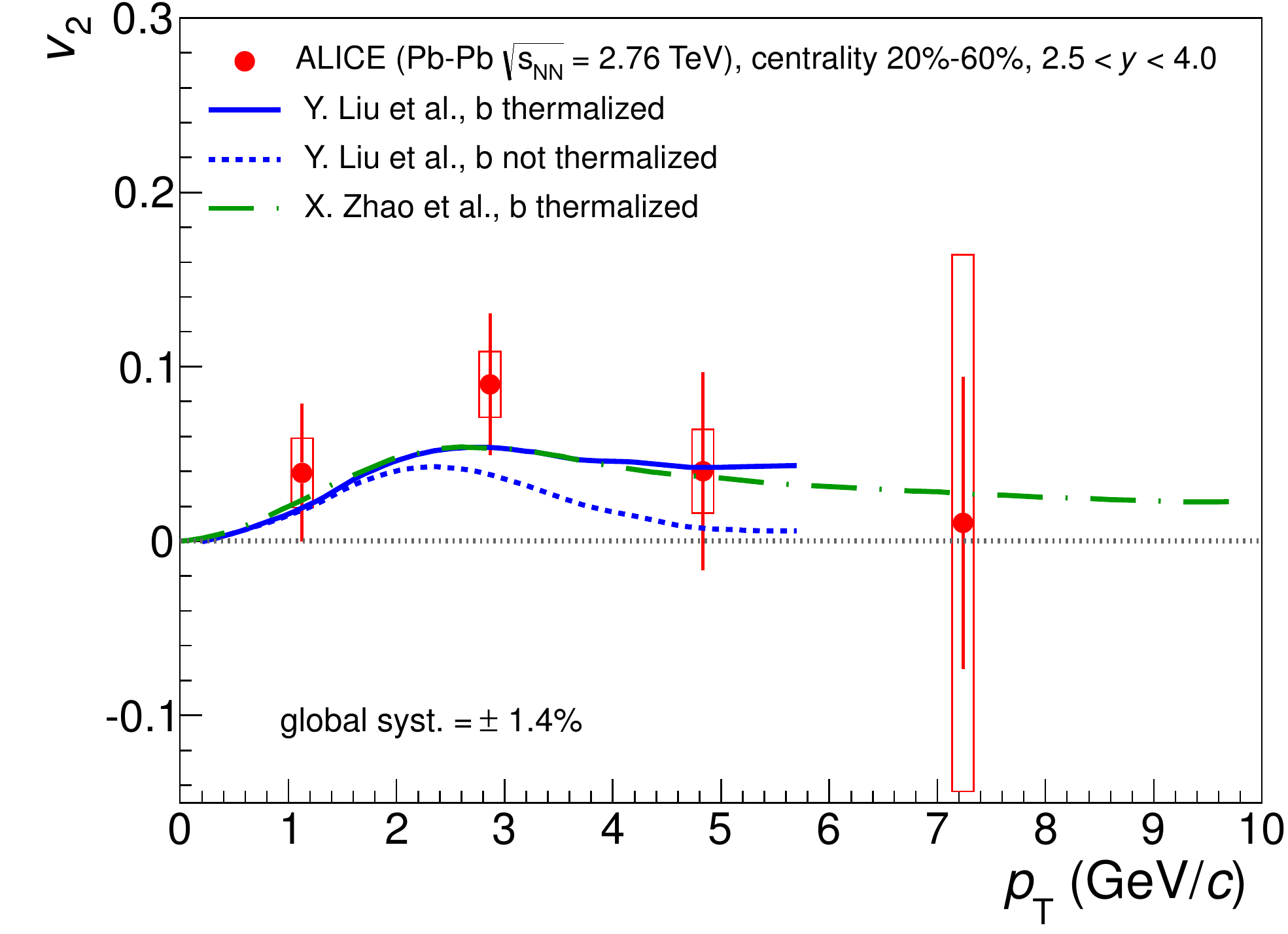}
\caption{(Color online) $J/\psi$ $v_{2}$ at forward rapidity (2.5$<$y$<$4.0) for non-central (20-60$\%$) Pb-Pb collisions at $\sqrt{s_{NN}}$  = 2.76 TeV~\cite{jpsi_lhc_data}. }
\label{jpsiv2_alice}
\end{center}
\end{figure}

ALICE collaboration also reported the measurement of the elliptic flow of $J/\psi$ in Pb-Pb collisions at $\sqrt{s_{NN}}$  = 2.76 TeV within the rapidity range 2.5$<$y$<$4.0~\cite{jpsi_lhc_data}. $J/\psi$ $v_{2}$ for non-central (20-60$\%$) Pb-Pb collisions at $\sqrt{s_{NN}}$  = 2.76 TeV is shown in Fig.~\ref{jpsiv2_alice}. Unlike RHIC, an indication of non-zero $J/\psi$ $v_{2}$ is observed with a maximum value of $v_{2}$ =  0.090 $\pm$ 0.041 (stat) $\pm$ 0.019 (syst) for  non-central (20-60$\%$) Pb-Pb collisions. Calculations from two transport models~\cite{jpsi_lhc_model1,jpsi_lhc_model2} are also shown for comparison. Transport model calculations that include a $J/\psi$ regeneration component (30$\%$) from deconfined charm quarks in the medium describes data very well. 

\section{Summary and Discussion}
In summary, the measurement of elliptic flow of heavy flavor can provide valuable information about the QGP medium. This article reviewed several important results from RHIC and LHC experiments and discussed their implications. In this review article, we have focused on measurements of heavy flavor $v_{2}$ as a function of $p_{T}$, collision centrality and energy carried out in RHIC and LHC experiments. We also discussed the comparison of these experimental measurements with available theoretical model predictions.  

Measurement of  azimuthal anisotropy of non-photonic electron is discussed at $\sqrt{s_{NN}}$ = 39, 62.4, 200 GeV and 2.76 TeV. NPE $v_{2}$ is consistent with zero at  $\sqrt{s_{NN}}$ = 39 and 62.4 GeV while it is non-zero $v_{2}$ at  $\sqrt{s_{NN}}$ = 200 GeV and 2.76 TeV. Large positive $v_{2}$ of NPE at low $p_{T}$  might indicate that charm quarks participate in the collective expansion of the dense and hot QGP. Elliptic flow of open charm $D$-meson measured by STAR and ALICE experiments are presented. A  non-zero positive flow has been observed at $\sqrt{s_{NN}}$ = 200 GeV and 2.76 TeV. Models that include recombination as mechanism of hadronization explain non-zero positive $v_{2}$ at low $pT$, however, a simultaneous description of $v_{2}$ and $R_{AA}$ is still an open issue. We also discussed the elliptic flow of $J/\psi$ measured at $\sqrt{s_{NN}}$ = 200 GeV and 2.76 TeV by STAR (at mid-rapidity) and ALICE (at forward rapidity).  $J/\psi$ flow is consistent with zero at RHIC, however, the measurement is statistically limited. A positive $J/\psi$ $v_{2}$ has been observed $\sqrt{s_{NN}}$ = 2.76 TeV.  Transport model calculations that include a $J/\psi$  regeneration component  from deconfined charm quarks in the medium describe data very well within current statistical uncertainties. 

A precise measurement of heavy flavor hadrons will provide information on fundamental properties of the medium, such as the transport coefficient, and hadronization mechanisms. As we discussed, current heavy flavor measurements are limited by statistics. In order to circumvent this, recent upgrades have been made in STAR with the introduction of the Heavy Flavor Tracker (HFT) and Muon Telescope Detector (MTD) and dedicated high statistics run in 2016. We hope to see the results from these soon. The ALICE experiment is also upgrading its detectors to pursue high precision measurements in the heavy flavor sector~\cite{alice_upgrade}. ALICE is upgrading Inner Tracking System (ITS), for better position and momentum resolution with a faster readout using frontier technologies. The upgraded readout data collection rate of ALICE is expected to increase by a factor of 100.  With the current set-up of ALICE, the flow analysis of $D_{s}$,  $\Lambda_{c}$ and $\Lambda_{b}$ is not accessible in Pb-Pb due to the limited statistics. Among all open charmed meson, $D_{s}$ has been considered as quantitative probe for charm quark hadronization. A comparison of $D_{s}$ and $D_{0}$ $R_{AA}$ and $v_{2}$ could be interesting to shed light on heavy quark dynamics. The new ITS is expected to allow for a precise measurement of the all $D$-mesons $v_{2}$ down to very low momentum~\cite{alice_upgrade}.  
 
\section{Conflict of Interests}
The authors declare that there is no conflict of interests regarding the publication of this paper.

\begin{acknowledgements}
This work is supported by DOE grant (Project No. 444025-HU- 206 22446) of Department of Physics and Astronomy, UCLA, USA.
\end{acknowledgements}

\end{document}